\documentclass[a4wide]{article}
\usepackage{amssymb}

\title{The shape dynamics description of gravity}
\author{Tim Koslowski\\Department of Mathematics and Statistics,\\ University of New Brunswick\\Fredericton, New Brunswick, E3B 5A3, Canada\\ \texttt{t.a.koslowski@gmail.com}}

\begin{document}
\maketitle

\begin{abstract}
 Classical gravity can be described as a relational dynamical system without ever appealing to spacetime or its geometry. This description is the so-called shape dynamics description of gravity. The existence of relational first principles from which the shape dynamics description of gravity can be derived is a motivation to consider shape dynamics (rather than GR) as the fundamental description of gravity. Adopting this point of view leads to the question: What is the role of spacetime in the shape dynamics description of gravity? This question contains many aspects: Compatibility of shape dynamics with the description of gravity in terms of spacetime geometry, the role of local Minkowski space, universality of spacetime geometry and the nature of quantum particles, which can no longer be assumed to be irreducible representations of the Poincar\'e group. In this contribution I derive effective spacetime structures by considering how matter fluctuations evolve along with shape dynamics. This evolution reveals an ``experienced spacetime geometry.'' This leads (in an idealized approximation) to local Minkowski space and causal relations. The small scale structure of the emergent geometric picture depends on the specific probes used to experience spacetime, which limits the applicability of effective spacetime to describe shape dynamics. I conclude with discussing the nature of quantum fluctuations (particles) in shape dynamics and how local Minkowski spacetime emerges from the evolution of quantum particles.
\end{abstract}

\maketitle

\section{Introduction}

A physical theory of the universe should describe the evolution of all physical objects, including those that we use to define units of length, time, frames of reference and all other dimensionful concepts. The fundamental description of the universe should thus be completely relational and dispense of all external reference structure. It turns out that not all physical systems can be described without external reference structure. I.e. relational considerations put restrictions on candidate theories for the dynamics of the universe. 

These relational restrictions on the theory of the universe can be illustrated in a simple toy model. Suppose a Newtonian universe with $N$ gravitating particles. If this is taken as a model of the entire universe then one should not appeal to any extraneous non--dynamical coordinate origin or reference frame or extraneous units of mass, length (scale) or time (duration). Independence of the dynamics from the choice of coordinate origin, reference frame and length- and time- scale requires that (1) the total linear momentum vanishes $\vec P=0$ (2) the total angular momentum vanishes $\vec J=0$ and that (3) the total energy vanishes $E=0$. If these conditions are imposed then one can describe its dynamics in relational terms, i.e. without having to appeal to any origin, scale, frame or notion of units. See \cite{BKMpot,BKMlong,BKMPRL} for a detailed description of this relational dynamical model. 

One might now worry that the current theory of the universe does not meet these strong relational criteria. However, it turns out that Einsteinian gravity can be described as a completely relational dynamical system, analogous to the relational $N$-body problem. This can be verified by going to the shape dynamics description of gravity \cite{GGK,GK1,ObsEquiv}, which I explain in the next section. This description of gravity can be derived from purely relational first principles \cite{CSplusV}. This is a motivation to consider relational foundations as fundamental to the description of the universe. The particular relational approach that underlies shape dynamics completely dispenses of spacetime \cite{RWR}. This means that the shape dynamics description changes the foundations gravity from spacetime geometry to a relational dynamical system. Spacetime and its geometry are, from the point of view of the shape dynamics description of gravity, merely effective concepts that describe physical experience. However, these effective descriptions may not exist as geometric objects in the strict mathematical sense.

The shape dynamics description of gravity thus allows one to consider the question whether ``the spacetime picture is an obstruction'' by investigating problems in gravity using shape dynamics. This approach may yield new insights in quantum gravity, the nature of spacetime singularities or the initial conditions for cosmology. But, before embarking on these ambitious projects, it is important to understand how spacetime geometry emerges from the shape dynamics description of gravity and what the limitations of this emergent spacetime picture are. I address this question in the remainder of this contribution by considering the evolution of weak matter fluctuations in the shape dynamics description of gravity in section \ref{sec:Classical}. This reveals that classical spacetime geometry can only be attained in an idealized simultaneous limit when matter fluctuations become arbitrarily weak (infinitesimal field strength) and modes with arbitrarily short wavelength are used. This limit is practically not attainable, so from the perspective of shape dynamics, spacetime geometry is an inherently fuzzy and scale-dependent object. Moreover, quantum mechanics teaches that it is in principle impossible to attain this limit, because particles described by arbitrarily short wavelength modes are expected to interact strongly with gravity. In section \ref{sec:Quantum} I introduce free quantum fields that evolve with shape dynamics as a first preparatory step in the investigation of these fundamental limitations.

\section{Gravity as a relational dynamical system}

The fastest introduction to the shape dynamics description of gravity is to introduce it as a gauge--unfixing of canonical General Relativity (GR) in York gauge. This hides the relational motivations for shape dynamics, but their detailed presentation goes beyond the scope of the present contribution. I refer the interested reader to the comprehensive tutorial \cite{FlavioTutorial} for a discussion of the underlaying relational ideas and the derivation of shape dynamics. The only relational restriction that I need for the following is that the Cauchy surface $\Sigma$ is compact and without boundary, so no boundary or asymptotic conditions introduce absolute structures that are in conflict with the relational foundations.

I start with the canonical description of GR \cite{DiracCanonical}, known as the ADM description of GR \cite{ADM}. The gravitational degrees of freedom are the spatial metric $g_{ab}$, its canonically conjugate momentum density $\pi^{ab}$, the lapse function $N$ and the shift vector $\xi^a$. The relation with the spacetime line element is
\begin{equation}
 ds^2=(g_{ab}\xi^a\xi^b-N^2)\,dt^2+2g_{ab}\xi^a\,dx^bdt+g_{ab}\,dx^adx^b.
\end{equation}
The shift vector field $\xi^a$ and the lapse function $N$ turn out to be Lagrange multipliers for the constraints:
\begin{equation}
 \begin{array}{rcl}
   H(\xi)&=&\int_\Sigma d^3x\,\left(\pi^{ab}(\mathcal L_\xi g)_{ab}+\pi^A(\mathcal L_\xi \phi)_A\right)\\
   S(N) &=&\int_\Sigma d^3x\,N\,\left(\frac{\pi^{ab}(g_{ac}g_{bd}-\frac 1 2 g_{ab}g_{cd})\pi^{cd}}{\sqrt{|g|}}\right.\\
             &&\left.-(R-2\Lambda)\sqrt{|g|}+\textrm{matter terms}\right),
 \end{array}
\end{equation}
where $\phi_A$ and $\pi^A$ stand collectively for matter fields and their canonically conjugate momentum densities. These constraints are initial value constraints and gauge generators, generating spatial diffeomorphisms and refoliations respectively. Time evolution is generated by the total Hamiltonian
\begin{equation}
 H_{\textrm{tot.}}=S(N)+H(\xi).
\end{equation}
This Hamiltonian is a linear combination of the constraints, which shows that the refoliation constraints entangle gauge invariance and time evolution inextricably\footnote{It is not possible to simply declare one of the $S(N)$ to be the physical Hamiltonian and all others to constraints, because the constraint algebra will not close in this case.}. 

A very versatile slicing condition (i.e. a gauge fixing condition for the refoliation generators $S(N)$) was introduced by York who expanded a proposal by Dirac \cite{DiracCMC}. This slicing condition is called the constant mean extrinsic curvature (CMC) condition
\begin{equation}
 \pi-\langle \pi \rangle\sqrt{|g|}=0,
\end{equation}
where $\pi:=g_{ab}\pi^{ab}$ and $\langle \pi \rangle:=\frac{\int_\Sigma \pi}{\int_\Sigma \sqrt{|g|}}$. Following York one can solve the solve scalar constraints uniquely (for details and exceptions see e.g. \cite{YorkOM,YIOM,Isenberg,IsenbergNester1,IsenbergNester2}) through a conformal factor $\Omega_o[g_{ab},\pi^{ab},\phi_A,\pi^A;x)$ that solves the Lichnerowicz--York equation
\begin{equation}\label{equ:LYE}
 \begin{array}{rcl}
   8\Delta\,\Omega& = &R\,\Omega +\left(\frac{\langle\pi\rangle^2}{6}-2\Lambda\right)\Omega^5-\frac{\sigma^a_b\sigma^b_a}{|g|}\Omega^{-7}\\
    &&+\,\,{\textrm{matter terms}},
 \end{array}
\end{equation}
where $\sigma^{ab}:=\pi^{ab}-\frac 1 3 \pi g^{ab}$ denotes the trace--free part of $\pi^{ab}$. It follows that the gauge-fixed Hamiltonian (for vanishing shift) is the on--shell value of the spatial volume
\begin{equation}
 H_{\textrm{York}}=\int_\Sigma d^3x\,\sqrt{|g|}\,\Omega_o^6[g_{ab},\pi^{ab},\phi_A,\pi^A].
\end{equation}
This Hamiltonian generates evolution in York time $\tau:=\frac 2 3 \langle\pi\rangle$, which turns out to be a monotonic evolution parameter for almost all physical data. 

This dynamical system can be gauge--unfixed to a system with spatial conformal symmetry. To see this, we first observe that the conformal transformation properties of the Lichnerowicz--York equation imply that $H_{\textrm{York}}$ is invariant under spatial conformal transformations that keep the total spatial volume $V=\int_\Sigma d^3x\,\sqrt{|g|}$ fixed. We then observe that the smeared gauge--fixing condition 
\begin{equation}
  Q(\rho):=\int_\Sigma\,\rho\,\left(\pi-\frac 3 2 \tau \sqrt{|g|}\right)
\end{equation}
has two properties: (1) it acts as the generator of conformal transformations of the spatial metric and the trace--free part of the metric momenta and (2) forms a closed constraint algebra with the spatial diffeomorphism generator. We can thus use the gauge--fixing condition to replace $\pi \to \frac 3 2 \tau \sqrt{|g|}$ such that the York Hamiltonian becomes $H_{SD}=H_{\textrm{York}}[g_{ab},\sigma^{ab}+\frac 3 2 \tau g^{ab}\sqrt{|g|},\phi_A,\pi^A]$ and from now on consider $\tau$ and $V$ no longer as expansion rate and spatial volume, but $\tau$ as an abstract evolution parameter and $V$ as a pure gauge quantity. This allows us to define a dynamical system with Hamiltonian $H_{SD}$ and gauge generators $ Q(\rho), \,H(\xi)$, which implement the relationally motivated invariances under spatial diffeomorphisms and spatial conformal transformations. The resulting relational theory will be referred to as shape dynamics.

This trading of gauge symmetries can also be obtained in a systematic way using a ``linking theory'' to facilitate gauge ``symmetry trading'' (see \cite{GK1}). This formalism shows explicitly that the physical predictions of shape dynamics are indistinguishable from GR whenever CMC gauge is admissible. However, shape dynamics and GR disagree in one important point: GR describes spacetimes and provides a full spacetime geometry, while shape dynamics implements relational principles and provides only the evolution of spatial conformal geometry. Neither spatial scale nor physical duration are provided by the shape dynamics description of gravity. These two concepts concepts are, form the perspective of shape dynamics, only effective descriptions of the evolution of matter fluctuations as I will show in the next section.

\section{Classical fields experience spacetime}\label{sec:Classical}

Albert Einstein himself noted in 1949 (see \cite{Stachel}) about the non--relational role of spacetime geometry in GR:\\ 
``{\it It is striking that the theory [...] introduces two kinds of physical things, i.e. (1) measuring rods and clocks, (2) all other things, e.g., the electromagnetic field, material point, etc. This, in a certain sense, is inconsistent; strictly speaking measuring rods and clocks would have to be represented as solutions of the basic equations... not, as it were, as theoretically self-sufficient entities.''}\\
It is my interpretation of this statement that one should not consider spacetime geometry as an entity independent of physical clocks and rods, rather spacetime geometry is an abstraction derived from the evolution of physical clocks and rods. This physical notion ``spacetime as a derived abstraction'' can be used to provide a notion of scale and duration in shape dynamics (see \cite{GKmatter} for the conceptual idea). In the following, I will consider fluctuations of matter fields to provide the physical clocks and rods (for a more detailed treatment see the forthcoming paper \cite{tim1}).

These fluctuations are supposed to be weak enough, so gravitational backreaction can be neglected for short enough time intervals. These two requirements (weakness of the fluctuation field strength and short evolution time) ensure that the physical clocks and rods do not significantly distort the geometry.

\subsection{Equations of motion for fluctuations}

I now consider a single component scalar field $\varphi$ (with canonically conjugate momentum density $\varpi$). I have to implement the no-backreaction requirements: (1)  to ensure that the scalar field is weak, I use a formal perturbation parameter $\epsilon$ and take the limit $\epsilon \to 0$ at the end of the calculation. (2) to ensure short evolution times, I consider only infinitesimal time steps. Hence, I consider the canonical equations of motion of $\sqrt{\epsilon}\,\varphi$ and $\sqrt{\epsilon}\,\varpi$. I assume that the action is even in $\varphi$, so the Hamiltonian admits a formal expansion $H_{SD}=H_0+\epsilon\,H_1+\sum_{n>1} \epsilon^n \,H_n$, where $H_0$ is independent of matter fluctuations and $H_1$ is quadratic in the matter fluctuations. It follows that
\begin{equation}
 \dot \varphi = \{\varphi,H_1\} +\mathcal O(\epsilon),\,\,\,\dot \varpi = \{\varpi,H_1\}+\mathcal O(\epsilon),
\end{equation}
so I have to derive only $H_1$ when considering the limit $\epsilon \to 0$. I will obtain $H_1$ from a formal expansion of the solution to the Lichnerowicz-York equation $\Omega=\Omega_0+\epsilon\,\Omega_1+\sum_{n>1} \epsilon^n\,\Omega_n$, where $\Omega_0$ is independent of of matter fluctuations and $\Omega_1$ is quadratic in them. Inserting this ansatz into the Lichnerowicz-York equation (\ref{equ:LYE}), in which each matter degree of freedom carries $\sqrt{\epsilon}$, and solving the first order in $\epsilon$ for $\Omega_1$ yields
\begin{equation}
 \Omega_1(x)=\int_\Sigma dy\,K(x,y)\,\frac{H_{matt}^{quad}}{\Omega_o\,\sqrt{g}}(y),
\end{equation}
where $K(x,y)$ is the integral kernel of the operator 
$$\left(8\Delta -\left(R+5(\frac 3 8 \tau^2-2\Lambda)\Omega_o^4+7\frac{\sigma^a_b\sigma^b_a}{|g|}\Omega_o^{-8}\right)\right)^{-1}$$
and where $H_{matt}^{quad}(y)$ refers to the quadratic part of the matter Hamilton density. In particular, for a scalar field we have:
\begin{equation}
  H_{matt}^{quad}(y)=\frac 1 2 \left(\frac{\varpi^2}{\sqrt{|\bar g|}}+(\bar g^{ab}\varphi_{,a}\varphi_{,b}+m^2\,\varphi^2)\sqrt{|\bar g|}\right),
\end{equation}
where I used the conformally corrected metric $\bar g_{ab}:=\Omega^4_o\,g_{ab}$ to simplify the notation. Inserting this solution into $H_{SD}$ and formally expanding the Hamiltonian yields the desired matter Hamiltonian
\begin{equation}\label{equ:matterHamiltonian}
 H_1=6 \int d^3x\,d^3y\,\sqrt{|g|}(x)\,\Omega_o^5(x)K(x,y)H^{quad}_{matt}(y)
\end{equation}
as the first order in $\epsilon$. 

Inspection of (\ref{equ:matterHamiltonian}) yields a formal lapse function
\begin{equation}
 N(x):=6\int d^3y\,\sqrt{|g|}(y)\Omega_o^5(y)\,K(y,x).
\end{equation}
It is the purpose of the next subsection to define physical clocks and rods to relate the formal lapse with a physical notion of duration.

\subsection{Experiencing spacetime geometry}

I will now use weak fluctuations of matter fields to define clocks, rods and equilocality. Experienced spacetime geometry is the geometry measured with these clocks and rods. My approach is still an idealization in the sense that I suppose that we are able to prepare arbitrarily accurate matter fluctuations and observe their evolution with arbitrary precision, but it is less of an idealization than presupposing a formal kinematic spacetime geometry. Formally, I assume an abstract topological space $\Sigma \times \mathbb R$ coordinatized by $(x\in\Sigma,\tau\in\mathbb R)$, which I use to keep track of the field strength of the matter fluctuations $\varphi(x)$ at York-time $\tau$.  

I start with the equations of motion for the matter fluctuations derived from (\ref{equ:matterHamiltonian}):
\begin{equation}\label{equ:matterEOM}
 \begin{array}{rcl}
   0&=&\ddot \varphi - N_o^2\,\Delta_{\Omega_o^4\,g}\,\varphi+N_o^2\,m^2\,\varphi\\
    &&+\,\dot\varphi\,\partial_\tau\left(\ln\frac{N_o}{\sqrt{|g|}\Omega_o^6}\right)+N_o\,N_{o,a}\,\Omega_o^{-4}g^{ab}\,\varphi_{,b},
 \end{array}
\end{equation}
where the first line contains the highest derivatives and the mass term. The first thing to observe from (\ref{equ:matterEOM}) is that the equations of motion do not exhibit any term that contains space and time derivatives of $\varphi$. This allows us to establish a notion of equilocality on $\Sigma$ in the sense that we do not have to implement a systematic shift in $\Sigma$ when evolving in $\tau$.

The second step is to deduce the light cone from observing how fluctuation wave packets propagate. The physical picture that I have in mind is to explore the light cone as the limit posed on the evolution of the support of fluctuations. For the infinitesimal time evolution of infinitesimally weak fluctuations, one can read this limit off of the principal symbol of the differential operator acting on $\varphi$ in (\ref{equ:matterEOM}), which is $\left(\partial_\tau^2 - N_o^2\,\Omega_o^{-4}\,g^{ab}\partial_a\partial_b\right)\varphi+\textrm{lower orders}=0$. We thus see, that a line element with the same light cone as the experienced spacetime geometry is
\begin{equation}\label{equ:ExpConfMetric}
 ds^2\,=\,-d\tau^2\,+\,\left(\frac{\Omega_o^2}{N_o}\right)^2\,g_{ab}\,dx^a\,dx^b
\end{equation}
This determines, by Malament's theorem \cite{Malament}, the spacetime geometry up to a spacetime conformal factor. 

The third step consists of determining this spacetime conformal factor. This is very simple if one has a massive scalar field, since the nonlinearity on the dispersion relation of a scalar fields sets a time scale. This time scale can be observed by preparing a superposition of local modes and observing the effect of the nonlinearity of the dispersion relation in the evolution of their interference pattern. By observing the effect of this nonlinearity, one does in effect define a unit of time $T_o^{-2}:=m^2$. The mathematical problem is thus to solve the local dispersion relation for $m^2$. 

To do this I have to explain what I mean with local modes and local dispersion relation at a point $x$: First, I choose a small enough neighborhood $U_x$ of $x$ so I can approximate the geometry fapp. as homogeneous. Next, I consider local exponential modes in $U_x$, i.e. in homogeneous local coordinates, I choose $\varphi(x,\tau)=e^{-i(\omega_x(k)\,\tau \pm \vec k\,\vec x)}$. Inserting these local modes into (\ref{equ:matterEOM}) and approximating all coefficients in (\ref{equ:matterEOM}) as homogeneous, one obtains the local dispersion relation of he form
\begin{equation}
 \omega^2_x(k)+u^o\,\omega_x(k)+N_o^2\,g_{ab}k^ak^b+u^a\,k_a+N_o^2\,m^2=0.
\end{equation}
We can thus use a local reference mode $k_o$, which has nodes only at the boundary of a reference interval and consider $2\,k_o$, $3\,k_o$ and $4\,k_o$, which are distinguished as modes in the reference interval that have nodes at the boundary and one, two or three nodes within. We see that the interference pattern changes as if the mass where $N_o\,m$ when coordinate time $\tau$ is used. We thus see that our unit of time implies that we have to reparametrize time using the lapse $N_o$, to implement the condition that the unit of time is everywhere given by $T_o$. The experienced line element is thus
\begin{equation}
  ds^2\,=-\,N_o^2\,d\tau^2\,+\,\Omega_o^4\,g_{ab}\,dx^a\,dx^b. 
\end{equation}
In the massless case, there is no scale provided by the dispersion relation, which implies that one can not determine the absolute scale of the spacetime conformal factor. This means that one can define a different physical reference process locally (at a point point $x$) as providing the units ticks of a clock. It is thus clear that there can not be any first order differential equation that describes any relation of reference clocks that have infinitesimal distance, since the first order differences are simply due to the synchronization of clocks. It is more complicated to show (see \cite{tim1}) that once a reference clock at $x$ has sent modes to synchronize six infinitesimally distant neighbor clocks, then the signals sent back from the six neighboring clocks can be used to derive a nontrivial relation between the running of the seven clocks. This relation is essentially the the CMC--lapse equation for the experienced lapse $N_e$. However, this equation is linear and thus determines the experienced lapse $N_e(x,\tau)$ only up to a factor $f(\tau)$. It is thus not surprising that $N_e=f(\tau)N_o$ with some unknown $f(\tau)$. The experienced line element is thus determined up to a global reparametrization of time
\begin{equation}
 ds^2=f^2(\tau)\left(-N_o^2\,d\tau^2+\Omega_o^4\,g_{ab}\,dx^a\,dx^b\right).
\end{equation}
If the system contains a bound system whose size can be used to define a time--independent reference length, then the factor $f(\tau)$ can be determined up to a global choice of units by requiring that the the time difference that a matter fluctuation requires to travel this reference length is independent of $\tau$.

\section{Local Minkowski space}

The most important feature of spacetime geometry is that it can be locally approximated by the geometry of Minkowski space. I will now show to what extent local Minkowski space can be experienced within the shape dynamics description of gravity.

\subsection{Local Minkowski data}

At any interior point $y$ in a spacetime, one can choose local Riemann normal coordinates $x^\mu$ in which $y$ lies at the coordinate origin and the line element locks like the standard Minkowski line element $ds^2=-(dx^o)^2+(d\vec x)^2+\mathcal O(x^2)$. Moreover, any CMC slice that contains $y$ can be boosted into the $x^o=0$ plane with deviations of $\mathcal O(x^2)$. Thus, for any initial data and any point $x$ at time $\tau$ it is possible to perform a diffeomorphism, such that the initial data for the shape dynamics description around this point is
\begin{equation}\label{equ:ApproxMinkowskiData}
 g_{ab}=\delta_{ab}+\mathcal O(x^2),\,\,\,\sigma^{ab}=\mathcal O(x),\,\,\,\textrm{(and matter data)}.
\end{equation}
Moreover, for any finite $\tau$ one can find a small enough neighborhood of $x$ and $\tau$, such that the effect of expansion is fapp. negligible. Thus, to explain how local Minkowski space comes out of the shape dynamics description of gravity, one has to experience the line element in a region that is governed by data (\ref{equ:ApproxMinkowskiData}). 

\subsection{York Hamiltonian for isolated subsystems}

Any boundary conditions, such as a region being asymptotically Minkowski space, violate the relational principles that form the basis of the shape dynamics description of gravity. The obvious solution to this problem is to consider boundary conditions as the effect of the rest of a fully relational universe has on a subsystem that is fapp. isolated. In particular, asymptotically Minkowski boundary conditions should arise when a very small subsystem is placed in a comparably very large but finite region of ``finite infinity,'' which itself is only a small part of a fully relational closed universe. 

The guiding physical situation that I have in mind is to consider our solar system (sun-Jupiter) as the subsystem, which is embedded in an fapp. empty region of finite infinity that is about $10^6$ times larger than the solar system. Moreover, the typical time scale for precesses in the solar system is a year, which is shorter than the time that it takes for light signals to cross the region of finite infinity. Moreover, it is perfectly conceivable that the solar system is embedded in a universe that is spatially closed at scales that are much larger than the Hubble horizon. Moreover, due to the relative smallness of the solar system and the shortness of the solar system's timescale compared to epochs of York time, one can safely ignore the effect of cosmological expansion. We are thus in a situation in which a typical subsystem process is fapp. described by a process that occurs in an asymptotically Minkowski spacetime although it really takes place in a fully relational universe. 

This situation shows how we should derive the approximate York Hamiltonian for a subsystem in fapp. asymptotically Minkowski space: We need to invent a model of the rest of the universe. The typical subsystem process will be unaffected by the detail of this model, since it takes more time for an influence of the rest of the universe to cross the region of finite infinity than for the typical process to unfold. We can thus create the simplest model and assume that the universe is compact without boundary, say its spatial topology is $S^3$, and that it is tessellated into a large number of copies of identical copies of the subsystem. We can thus simply derive the dependence of the York Hamiltonian on the subsystem data by calculating the on--shell value of the spatial volume using this model for the rest of the universe. 

This particular model for the rest of the universe simplifies calculations because one only needs to consider only one tile of the tessellation; the York Hamiltonian is just the number of tiles times the on--shell value of the volume of one tile. We consider now a subsystem at the origin of the tile that is surrounded by a region of finite infinity, which is in Minkowski vacuum
\begin{equation}\label{equ:MinkowsloID}
 g_{ab}=\delta_{ab},\,\,\,\sigma^{ab}=0,\,\,\,\,\textrm{(and vanishing matter data)}.
\end{equation}
The Lichnerowicz--York equation in the region of finite infinity is $\Delta\,\Omega=0$. The coordinate radius $\psi_o$ of region of finite infinity is much larger than the radius $\psi_f$ of the subsystem. The ratio $L=\psi_o/\psi_f$ can thus be assumed to be fapp. infinite. This implies the boundary condition $\lim_{r\to \infty \infty} \Omega(r)=const.$. We can define our units s.t. $const.=1$. The solution to the Lichnerowicz--York equation is thus
\begin{equation}\label{equ:LYsolution}
 \Omega(r,\theta,\phi)=1+\frac{a_o^-}{r}+\sum_{l=1}^\infty r^{-(l+1)}\sum_{m=-l}^{l} a^+_{lm}\,Y_{lm}(\theta,\phi),
\end{equation}
where $Y_{lm}$ denote the spherical harmonics. This solution depends on the the subsystem initial data through the coefficients $a^\pm_{lm}$. The on--shell value of the volume of the tile is
\begin{equation}
 V_{tile}=L^3\,C+L^2\,\frac{1}{2\pi}\int_{\partial\textrm{tile}} d^2x^a\,\Omega_{,a}(x)+\mathcal O(L),
\end{equation}
where $C$ is independent of subsystem data. The York Hamiltonian is proportional to $V_{tile}$. It generates evolution in York time. Before taking the limit $L\to\infty$ (which implements $L=$fapp.$\infty$), we have to reparametrize time to ``solar system'' time. To find out what this reparametrization has to be, we consider a subsystem that consists of a weak scalar field fluctuation for which we can derive the Hamiltonian as described in the previous section and then evolve the fluctuation with the Hamiltonian $V_{tile}$. This gives:
\begin{equation}\label{equ:MinkowskiEOM}
 \ddot \phi\,-\,c^2\,L^4\,\Delta\,\phi=0,
\end{equation}
where $c$ is a constant of $\mathcal O(1)$. The reparametrization form York--time $\tau$ to solar--system time $t$ is thus
\begin{equation}
 t:=L^2\,\tau.
\end{equation}
The effective Hamiltonian for the subsystem is thus
\begin{equation}\label{equ:MinkowskiHamiltonian}
 H_{sub}=\frac{1}{2\pi}\int_{\partial\textrm{tile}} d^2x^a\,\Omega_{,a}(x)+\mathcal O(L^{-1}).
\end{equation}
This expression coincides with the ADM energy in quasi--isotropic gauge in the limit $L\to\infty$.
 
\subsection{Experiencing Minkowski geometry}

We can now use the effective subsystem Hamiltonian (\ref{equ:MinkowskiHamiltonian}) to explore what geometry is seen by a weak scalar fluctuation in a region with Minkowski initial data (using the method described in the previous section). It is clear from equation (\ref{equ:MinkowskiEOM}) that the experienced geometry is conformal to Minkowski space, since the equation is isotropic in space and the speed of light is homogeneous in space. Moreover, using a massive scalar fluctuation, we find that the unit duration $N_o$ is homogeneous as well, so the spacetime geometry experienced by a weak scalar fluctuation is described by a Minkwoski line element.

\subsection{Universal geometry and universal coupling}

We have just seen that (1) the initial canonical data is given by approximate Minkowski data (\ref{equ:ApproxMinkowskiData}) and (2) that the geometry experienced by a weak scalar field is described by a Minkowski line element when the fluctuation is evolved with Minkowski initial data. This shows that ``local Minkowski space'' of GR emerges in the shape dynamics description of gravity. However, we have only shown this for one species of fields. This leads to the question of whether geometry is universal, i.e. whether all particles experience the same spacetime geometry. 

The fact that all species experience the same spacetime geometry is a consequence of the formal property that matter couples universally to the spacetime metric of the GR description of gravity (e.g. by minimal coupling). The shape dynamics description does not possess a formal notion of universal coupling. Hence it is not a question whether all species experience the same geometry. Rather, the postulate that the experienced geometry is independent of the type of fluctuation that is used to experience geometry constitutes the shape dynamics analogue of GR's universal coupling.

\subsection{Limitations of spacetime geometry}

So far I have laid out the argument that matter fluctuations that evolve together with the shape dynamics description of gravity experience a spacetime geometry which coincides locally with the spacetime geometry of the GR description of gravity. However, unlike GR, there is no spacetime geometry in the shape dynamics description. Spacetime geometry emerges only when idealized clocks and rods are evolved. These idealizations are in particular: (1) clocks and rods are infinitely light, s.t. they do not distort the geometry they are probing, i.e. there is no gravitational back reaction. (2) the clocks and rods are infinitely localized, i.e. the modes used to probe geometry have arbitrarily short wave--length. (3) the clocks and rods can be prepared with infinitely high precision. (4) the states of the clocks and rods can be read with infinitely high frequency.

None of these idealizations can be met by actual physical clocks and rods. This imposes limitations on the applicability of experienced spacetime geometry as an effective description of a shape dynamics solution. Spacetime geometry is an idealized abstraction. If one e.g. appreciates the fact that any physical clock has to be built from finite wavelength modes, then one obtains a scale--dependent notion of experienced geometry already at the classical level. Moreover, one expects that quantum uncertainty of matter fields, which obstructs light clocks from being arbitrarily localized, poses fundamental limitations on the notion of geometry. This motivates the following section in which I consider the evolution of quantum fields coupled to shape dynamics.

\section{Quantum fields in shape dynamics}\label{sec:Quantum}

I will now consider the evolution of fluctuations of free quantum fields along with shape dynamics. In particular, I consider the shape dynamics analogue of a free QFT in curved spacetime (for a motivation of this approach and a more detailed treatment see the forthcoming paper \cite{tim2}).

\subsection{Symmetries of isolated subsystems}

The York Hamiltonian of a subsystem is by construction invariant under spatial diffeomorphisms and spatial conformal transformations. However, Minkowski data in the region of finite infinity is only invariant under those transformations that leave Minkowski data invariant. These are the conformal Killing fields of the 3--dimensional Euclidean metric. The generators of these transformations are conserved by the shape dynamics evolution. These conserved quantities are experienced as conserved charges of the subsystem by Minkowski observers in the region of finite infinity. The standard basis for the conformal Killing fields of the 3--dimensional Euclidean metric is
\begin{equation}
 \begin{array}{rclcrcl}
   \xi^T_a &=& \partial_{x^a}&\quad&
   \xi^R_a &=& \epsilon_{abc}\, x^b \,\partial_{x^c}\\
   \xi^D   &=& r\,\partial_r&&
   \xi^S_a &=& 2 x^a\,x^b\,\partial_{x^b}-r^2 \partial_{x^a},
 \end{array}
\end{equation}
which generate translations, rotations, dilatations and special conformal transformations respectively. The associated conserved charges are the values of the Poisson generators of the automorphisms of the Minkowski data
\begin{equation}
  C_A=H(\xi_A)-\frac 2 3 Q(\textrm{div}(\xi_A)).
\end{equation}
The Poisson--algebra of these charges is the Lie--algebra $C(3)$ of the conformal group in three dimensions, which coincides with the Lie--algebra of the de Sitter group $SO(4,1)$.

The leading order in a large--$L$-expansion of value of the $C_A$ can be calculated using the on--shell value of the metric momenta $\pi^{ab}=\frac 1 2\,\tau\,g^{ab}\sqrt{g}$ as well as the expression (\ref{equ:LYsolution}) for $\Omega$ in the region of finite infinity. It follows from my ansatz that the leading order in $L$ of the linear and angular momentum charge will vanish. Upon a time reparametrization from York time to solar system time (as in the previous section), which introduces a factor $L^{-2}$, I find that the dilatation- and special conformal charges become
\begin{equation}
 \begin{array}{lcl}
   D\,&=& -12\,\tau\,\pi\,a_0^- + \mathcal O(L^{-1})\\
   S_z&=& -12\,\tau\,\sqrt{\frac \pi 3}\,a_{1,0}^+ + \mathcal O(L^{-1})
 \end{array}
\end{equation}
where I assumed without loss of generality that the $z$-direction of my frame is aligned with the direction of the special conformal charge. We see that the dilatation charge is proportional to $\tau$ times the subsystem Hamiltonian $H_{\textrm{sub}}=12\,\pi\,a_0^{-}+\textrm{const.}+\mathcal O(L^{-1})$. Moreover, for the purpose of finding the charge algebra in the limit $L\to\infty$, we can restrict ourselves to studying the representation of $C(3)$ as generators of spatial conformal isometries, and identify the generator of time evolution $-\tau\,D$.

\subsection{Canonical quantization}

The most natural\footnote{One can motivate this natural quantization by considering a saddle--point approximation of a formal (but ill--defined) path--integral of a shape dynamics+matter system.} quantization of fluctuations of quantum fields that evolve with shape dynamics is based on quantizing the classical charge--algebra of the $C_A$ and to identify the generator of time evolution with the $1/t$ times the dilatation generator. Concretely, I propose the following canonical quantization

\begin{enumerate}
 \item The Hilbert space $\mathcal H$ of fluctuations carries a unitary representation of $C(3)$.
 \item $\mathcal H$ contains a cyclic $C(3)$-invariant vacuum state $\left|\Omega_\omega\right\rangle$.
 \item\label{item:PosEnergy} The spectrum of the dilatation operator $\hat D$ in $C(3)$ is non-positive after reparametrization to solar system time.
 \item Time evolution is generated by $-\tau\,\hat D$.
\end{enumerate}
Notice that these axioms differ significantly from the axioms of spacetime conformal QFTs.

The first postulate implies that $\mathcal H$ is a sum of unitary irreducible representation (UIR) of $C(3)$. This is the starting point for canonical quantization, but to proceed practically I have to assume standard canonical second quantization:
\begin{enumerate}
 \setcounter{enumi}{4}
 \item Elementary excitations $|\vec k,s\rangle=a^*_{\vec k,s}|\Omega\rangle$ are generated by the action of creation operators $a^*_{\vec k,s}$, which satisfy canonical commutation relations $[a^*_{\vec k,s},a_{\vec k^\prime,s^\prime}]_{\pm}=\delta_{(\vec k,s),(\vec k^\prime,s^\prime)}$ with annihilation operators $a_{\vec k,s}$.
 \item The vacuum is empty $a_{\vec k,s}|\Omega\rangle=0$.
\end{enumerate}
Following \cite{Thomas}, the UIR of $C(3)$ can be expressed as a combination of the $SO(4)$ generators $J^i_a=J_a\pm(P_a+(2i-3)S_a)$, where $i=1,2$, with the conformal generators $A_\pm=-D\pm i\,(P_3-S_3)$ and $B_\pm=(P_1-S_1)\pm\,i(P_2-S_2)$. The $J^\pm_a$ act as angular momentum operators on the $1$ and $2$ components of the orthonormal basis $\langle j_1m_1,j_2m_2|$, while the conformal generators act as 
{\small $$
 \begin{array}{l}
   \langle j_1m_1,j_2m_2| A^{\pm}=\\
   \quad\mp\langle j_1^-m_1^\mp,j_2^- m_2^\pm| f_+(j_1,j_2)\sqrt{(j_1\pm m_1)(j_2\mp m_2)}\\
   \quad+\langle j_1^+m_1^\mp,j_2^- m_2^\pm| f_-(j_1,j_2)\sqrt{(j_1\mp_1+1)(j_2\mp m_2)}\\
   \quad+\langle j_1^-m_1^\mp,j_2^+ m_2^\pm| \overline{f_-(j_1^-,j_2^+)}\sqrt{(j_1\pm m_1)(j_2\pm m_2+1)}\\
   \quad\mp\langle j_1^+m_1^\mp,j_2^+ m_2^\pm| \overline{f_+(j_1^+,j_2^+)}\sqrt{(j_1\mp m_1+1)(j_2\pm m_2+1)},
 \end{array}
$$}
here $j_i^+:=j_i+\frac 1 2$ and $j_i^-,m_i^+,m_i^-$ analogously. The matrix elements for $B^{\pm}$ are obtained through an $SO(4)$-rotation of the matrix elements of $A^{\pm}$, where the $f_\pm(j_1,j_2)$ satisfy $|f_\pm(j_1,j_2)|^2=\frac{(j_1\pm j_2-p)(j_1\pm j_2+p+1)(j_1\pm j_2-q)(j_1\pm j_2+q+1)}{4j_1 j_2(2j_1+1)(2j_2+1)}$. The arbitrary constants $q,p$ are related to the eigenvalues of the two Casimir eigenvalues by $Q_2=-(p(p+1)+(q-1)(q+2))$ and $Q_4=-pq(p+1)(q+1)$. Unitarity implies that the possible values of $q,p$ come in three series: the {\it principle} series with $p\in\mathbb N_0, q=\frac 1 2+i \nu; \nu\ge 0$ and $p\in \frac 1 2 \mathbb N,q=\frac 1 2+i \nu;\nu>0$, the {\it complimentary} series with $p=0,q\in(-1, \frac 1 2)\cup(\frac 1 2,2)$ and $p\in \mathbb N,q\in(0,\frac 1 2)\cup(\frac 1 2,1)$ and the {\it discrete} series with $p\in\mathbb N,q=0,1,...,p$ and $p\in \frac 1 2 \mathbb N, q=\frac 1 2,\frac 3 2,...,p$. The quadratic Casimir is
\begin{equation}\label{equ:QuadCasimir}
 Q_2=D^2-\vec J^2-2(\vec P.\vec K+\vec K.\vec P)
\end{equation}
The quadratic part of the Hamiltonian $H=-D$ (in units with $\tau=1$) is determined by the representation of the elementary excitations, so  after normalizing $H$ is:
\begin{equation}\label{equ:Hamiltonian}
 H=-\sum_{b,b^\prime}\,a^*_b\,(D)_{b,b^\prime}\,a_{b^\prime}+\mathcal O(a^3),
\end{equation}
where $b,b^\prime$ extend over the entire basis and $D_{b,b^\prime}$ denotes the respective matrix element of $D$. Conventionally, one calls $H$ {\it free} if $\mathcal O(a^3)$ vanishes. 

\subsection{Free quantum fields in shape dynamics}

The scalar sector corresponds to principle series UIR with $p=0,q=\frac 1 2+i \nu$, where $\nu\ge 0$. This implies $Q_2=-(p(p+1)+(q-1)(q+2))\,\mathbb I$, which can be solved for $D$
\begin{equation}\label{equ:freeD}
 D_\pm=\pm\sqrt{J^2+2(\vec P.\vec S+\vec S.\vec P)+\left(\frac 9 4+\nu^2\right)\,\mathbb I},
\end{equation}
where I used a spectral square root. The Hamiltonian for a free scalar theory is thus
\begin{equation}
 H=\sum_{b,b^\prime}a^*_b\,(D_+)_{b,b^\prime}a_{b^\prime}.
\end{equation}

{\it Towards Minkowski spacetime:} It is well--understood that the Poincar\'e group can be obtained as a Wigner--In\"on\"u contraction of the de Sitter group. This is usually done by rescaling $D,Y^a=P^a-S^a$ with a contraction parameter $\lambda^{-2}$ and taking the limit $\lambda\to\infty$ while holding the mass $m:=\frac{\nu}{\lambda}$ fixed\footnote{The reparametrization from York--time $\tau$ to solar system time, i.e. the limit $L\to\infty$ above, has a similar effect. The difference with the standard contraction is that $D$ and $S^a$, rather than $D$ and $Y^a$, are contracted.}. For free fields the standard contraction transforms $D_{+}$ into
\begin{equation}
 D_+=\lim_{L\to\infty}\sqrt{L^{-2}\vec J^2+2(\vec X^2-\vec Y^2)+m^2\,\mathbb I},
\end{equation}
where $X^a=P^a+S^a$. This reduces to the Poincar\'e free scalar Hamiltonian $H=\sqrt{\vec{p}^2+m^2\,\mathbb I}$ when the Minkowski space momenta $\vec p$ is identified with $\vec X$. 

\section{Conclusions}

The shape dynamics description of gravity does not provide any a priori notion of spacetime geometry. This is due to the derivation of shape dynamics from relational first principles which do not presuppose the existence of spacetime geometry. This poses the question: ``What is the physical role of spacetime geometry in the shape dynamics description of gravity?''

I provided a partial answer to this question by showing that spacetime geometry is an abstraction that descries how idealized matter fluctuations evolve with the shape dynamics description of gravity. This (idealized) experienced spacetime geometry coincides locally\footnote{It is possible that the experienced geometry is, at a global level, different from the spacetime geometry of GR solutions.} with the spacetime picture provided by solutions of GR. In particular, I showed that the CMC--lapse and the Lichnerowicz--York conformal factor appear as the experienced duration and spatial scale respectively. I also showed why a weak fluctuation will evolve as if it was in local Minkowski space and that the question of universality of the experienced geometry leads to a notion of universal coupling that is applicable to shape dynamics. 

I emphasized the fact that the experienced geometry was derived in an idealization that is physically unattainable. This shows that the description of a shape dynamics solution through spacetime geometry has an inherently limited domain of applicability. Ultimately, one has to appreciate that all physical clocks and rods are finite physical systems, so the idealization in which GR's spacetime geometry is derived can never be achieved. In this way, one is lead to expect a scale--dependent notion of spacetime geometry.

In an effort to find a fundamental limitation of the spacetime description due to quantum uncertainty, I considered the evolution of free quantum fields along with shape dynamics. As a warm up for this problem I have shown how a locally Poincar\'e--invariant free QFT emerges from quantizing matter fluctuations in shape dynamics in a nontrivial way. This provides the framework for a systematic study of the question: ``What is the spacetime geometry that is experienced through quantum matter?'' 

\subsection*{Achnowledgements}

I thank the organizers of Theory Canada 9 for the invitation to give this talk. I have to give particular thanks to Julian Barbour and to my closest collaborators Henrique Gomes and Flavio Mercati. I also thank Sean Gryb, Lee Smolin and Timothy Budd as well as Viqar Husain, Niayesh Afshordi, Edward Anderson, Steffen Gielen, Hans Westman, Vasudev Shyam, Niall O'Murchadha and Phillipp H\"ohn. This work was in part supported by the Natural Sciences and Engineering Research Council of Canada through a grant to the University of New Brunswick and by the Foundational Questions Institute.

\end{document}